\documentclass[a4paper,11pt]{article}
\usepackage{adjustbox}
\usepackage{xcolor}
\usepackage{amsmath,latexsym}

\usepackage{adjustbox}

\begin{document}
\begin{center}
{\huge Deriving Locality, Gravity as Spontaneous Breaking of Diffeomorphism
  Symmetry}\\
\vspace{7mm}
{\bf H.B. Nielsen\footnote{Speaker at the  Work Shop
    ``What comes Beyond the Standard Models'' in Bled, 1923},\\
  Niels Bohr Institut,
(et al?)}\\
{Bled , July or Noveber  , 2023}
\end{center}

\begin{abstract}
  Looking for ideas for a fundamental physics that could almost work by itself,
  not making but mild assumtions, we are inspried by the earlier work by
  Astri Kleppe and myself. We showed that assuming diffeomorphism symmetry
  in addition to only ``mild assumptions'' we could obtain obtain effectively
  a local Lagrangian density,{\bf without assuming that the action were at all
    local!} However, we point out in the present article, that unless there is a
  spontaneous breakdown of the full diffeoemorphism invariance by
  gravitational like fields having non-zero vacuum expectation values, there
  would be no propagation possible, in other words the resulting effective
  local theory would be superlocal, so no communication between the different
  space-time points. We suggest that having a projective space or some similar
  space with some more structure, it could be in practice enough to get
  the effective locallity, so the full diffeomorphism symmetry might not not be
  really needed. Finally we found a phenomenological feature of the CMB
  (=Cosmic Microwave Backgrouns Radiation)
  low $l$ spherical harmonic expansion of the CMB fluctuations which we
  interpret as a sign of space-time in the background in which space-time is
  imbedded is a projective space, rather than just a manifold.
  \end{abstract}

\section{Introduction}

In the search for a theory beyond the physics we know today we have long
attempted ``Random Dynamics'', which consists of asking:

Are there some laws of nature that {\bf can be derived} from
  other ones
  in some (e.g. low energy-) limit ? If so we should leave the derivable law
  out of the fundamental model, and assume the ones needed in derivation as
  {\bf a
  step more fundamental}.
\vspace{1mm}

The example today: {\bf Locality $\Leftarrow$ Diffeomorphism symmetry}
(this derivation Astri Kleppe and myself \cite{locality} already claim to have
performed/ proved) 
\vspace{1mm}

So: Assume that {\bf diffeomorphism symmetry} or something similar
- e.g. projective geometry symmetry, or simplectomorphic symmetry -
is a very fundamental principle, while {\bf we do not assume locality
  as fundamental} at the same level.
\vspace{1mm}

In the future  we shall of course attempt {\bf also to derive diffeomorphism
  symmetry}
from something
else, and that might e.g. be by assuming some ``large amount'' of symmetry
acting on a space. Actually Masao Ninomiya and myself have stressed, that an
infinite space acted upon in an sharply 3-transitive (see these concepts
``sharply 3-transitive'' below in section \ref{Trans}) way is already close to
be the projective line\cite{projectiveline}.

\subsection{Philosophic Speculative Introduction, Plan}

Let us argue a bit looking at the present work as seeking a theory for
gravity, behind or beyond gravity:

  \begin{itemize}
  \item Introduction (we do not know what is behind gravity)
    
  \item Argument for a geometry with local scale and projective symmetry:

    \begin{itemize}
    \item Astri Kleppe and I could derive {\bf locality} of the
      effectively resulting action. We take it that since we can get
      locality out from starting from a symmetry postulate that has the wole
      symmetry of the manifold, {\bf without putting in explicitely locality},
      then it means that we can avoid the extra assumption, i.e. we have
      asimplification, if we assume such a symmetry like the whole
      diffeomorphism symmetry.
      
    \item Now we Only get gravity after a background field is assumed,
      like $g^{\mu\nu}(x)$. Actually unless we have such a field non-zero in
      vacuum, we get no propagation, i.e. the different space time points
      do not get connected. So you make say: We need as a very abstract
      need for physical theory, that there is connection between the different
      space time events. Then we must have some field like $g^{\mu\nu}(x)$ or
        some vierbeins or the like that can do the same job of introducing
        propagations of the field through space time. This is not really a
        derivation of gravity, but it is a little bit in that direction in
        as far as this argument present a {\bf need} for gravity.
        In usual theories likegenrel relativity one could say that we just have
        introduced gravity because of phenomelogical need, one simply has
        known for long effects of gravity, and thus we need it. Here I seek
        to say: we need something less specific, propagation of particles, and
        then because we have for other reasons(the beauty of deriving
        locality) assumed too much symmetry, we have come to need some fields,
        which of the most obvious type looks just like gravity fields.
        So we came to assume too much symmetry so that gravity could not be
        avoided for {\bf other than simply phenomelogical reasons of seeing
          gravitational forces}.

        In this point of view the gravity fields represent the fields
        giving the certain needed spontaneous break down of the too much
        assumed symmetry.

        This might make us think that it is less fundamental, but of course
        it of something from which it can be constructed has to be in the
        theory.
        Psykologically this way of looking at it might make or a bit less keen
        on looking at the effects of the fundamental theory, we look for,
        expected from gravity, because after all the gravity was just the
        presumably composite field breaking the too much assumed symmetry.

        That is to say we could use this way of thinking as an encouragement
        not to look for that we should have to bother with all the terrible
        topologically complicated space times, which are almost unavoidable in
        quantum gravity. They are in so huge amounts that they would be
        pretty hard to treat mathetically. 
    \end{itemize}
  \item Projective geoemetry (a possibility beyond gravity)

    Rather we might take the here put forward suggestion of looking at
    gravity as being imbedded into the manifold. The manifold is there at  the
    most fundamental level, at first we could say. But then we may get the
    hope that the terrible topological forms of manifolds might be avoided by
    modifying a bit the theory by replacing the manifold by a topologically
    less terrible structure with sufficiently similar symmetris that we could
    at least approximately still derive the locality principle. A proposal
    of a very nice structure behaving similarly to the manifold but with
    much simpler topological form is the {\bf projective spaces.}

    So the speculative model which we might propose here is that we
    live in back ground space time which is a projective space.
    The structure of such a space time is so simple that seen form small scale
    it is like a flat space, and thus has the promissing feature af providing
    a kind of explanation for how flat space time is in practice.

  \item A further hope brought by the projective space idea, is that
    there is hope to characterize the projective space by defining it by means
    of its symmetry properties. Indeed Masao Ninomiya and myself recently
    brought attention to that the projective line (= the projective pace
    of dimension 1) were close to come out by assuming a space to be
    sharply 3-transitively transformed under a group. Just assumming such
    3-transitivity you begin to see structures like the field (the real field
    say) without putting in the concept of a field directly. remeber that
    the manifold is defined by means of coordinates, which are differentable
    with respect to each other, and thus you define manifolds mathematically
    in a way already using the field-concept as wellknown. If the dream
    could come through of defining the background space time (now assumed to
    be a projective space time) without explicitely putting in our real numbers,
    but rather getting it out somewhat similar to studies of $3$-transitive
    actions of groups, then it would be very nice and suggestive on us being
    on the right track for the fundamental theory, because we could then
    claim: we did not even put in our real numbers, no they came out instead!

    Even though this requirement 3-transitivity only leads to
    `almost fields'' and not completely to the real fields as one is
    accustomed to use in the physical geometry, it is till much better than
    to have to put the fields in completely from ouside.
    
  \item A phenomenological support of projective goemetry.

    Then what could in the long run if it works out really support the idea
    of a background projective space would be if what we shall present below in
    section \ref{sPheno} is really working out and we have a phenomelogical
    observational indication that we indeed live in a world imbedded in a
    projective space time.

  \item Conclusion:

    A very optimistic Random Dynamics dream, might look
    like this: Almost whatever a very complicated mathematical structure
    would be like, it would if it is very big unavoidably have some similarity
    of some parts with some other parts. Such a similarity
    - presumably approximate only - would naturally be expressed by some
    approximate {\bf symmetry}, which of course in the mathematical
    language means that there is some group $G$ acting on the
    compicated structure being the world, say $X$, just as talked about
    in section \ref{Trans}. Then it is needed to find out which properties
    of such a system a group acting on a set/structure is most likely to be
    the type relevant for such an attempt of a theory of everything.

    With some empirical support I and Don Bennet found what characterizes
    the Standard Model group $S(U(2)\times U(3))$ (which is gauge group in
    O’Raifeartaigh.s sense \cite{ORaifeartaigh} (Group Structure of Gauge
    theories,University
    Press Cambridge (1986)), namely that the representation of the
    acting gauge group could be found to be essentially in volume the smallest
    possible compared to a volume constructed for the group
    itself\cite{Smallrep,Smallrep2,SmallrepDon, Smallrep3}. Taking this to
    say `` The group shall be so large compared to the set7structure it acts
    on as possible ( in Nature of fundamental physics)'', and noting that
    that the $n$ in the (sharp) $n$-transitivity of a group action roughly
    means that the group is the $n$th cross product power of the
    set $X$ on which it acts. In fact you can say that the group has been
    brought in correpondance to the cross product $X\times X \times \cdots
    \times X$ ( with n factors $X$=). Really a manifold or a projective space
    are spaces with relativly a lot of symmetries, so there is good hope
    to find that such a big group compared to the object acted upon
    could favor just imbedding spaces, we suggest in this article.
    But then since gravity is needed for getting propagation and the same
    big group compared to the representaiion or almost the same the space
    acted upon, then the Standard model gauge group could also come from such
    principle. Taking it that the fermion represetations and the Higgs
    representation in the Standard Model are indeed among the {\bf smaalest}
    faithful ones, there isnot much in our present knowledge about the
    physical laws, which would not be almost unavoidable the here sugested
    system.
    \end{itemize}

\subsection{Listing of Sections}

In the following section \ref{theorem} we shall present and discuss  the
mentioned theorem
of Astri Kleppes and mine, in which we get locality without putting it in,
while
the detailed proof of this theorem will be postponed to
section \ref{Locderivation}
.
Next in section \ref{propagation} we put forward the problem, that we cannot
get a world with propagation of fields/particles without a spontaneous breakdown
by having a non-zero $g^{\mu\nu}$-field (with upper indices) in vacuum. In the
subsection \ref{dreduction} we give a new way of making dimensional
reduction in the
spirit of the world being imbedded into the manifold (this is the manifold
always used in gerel Relativity)   or some projective space
which is similar to the manifold. In the subsection\ref{resume} we again
review the point of the gravity being needed for propagation.

In section \ref{Locderivation} we give the real proof of our derivation
of the principle of locality together of course with the statement of the
``mild'' assumptions, such as the analyticity - as a function{\bf al} - of
the otherwise so general action, that it is {\bf not} by assumption local.

Now it is often the most interesting about Random Dynamics derivetions, that
they do not succed completely and also the derivation of locality is only
partly succesful. In section \ref{sMPP} we thus tell that one of the results
from this not quite succesfullness of the locality derivation is that we
obtain an idea to derive (with in addition only very ``mild'' assumptions) an
old postulate of ours called `` Multiple Point Criticallity Principle'' (=MPP).

In the next section \ref{Trans} we  review the mathematical concept of
a group action acting $n$-transitively.

In section \ref{sRepeat} we suggest that one should excercise finding the
effective action - the now local one - which obeys the symmetries in our model
left over, so that we can give at least an idea about have one by a bit more
work might see that essentially the usual einstein general relativity comes out
of the model, in this article we have mainly left this for the reader, but we
hope to come through in another article.

The next section \ref{sprojective} is then mainly a review of projevtive
geometry, which we consider a promissing candidate for a space to replace
the full manifold. It may not lead to quite as perfect locality as the manifold,
but after all phenomenological quantum field theories, as we know them and
use them, have deviations from locallity at very short distencies, so if we get
that in our model, it might be an advantage. Indeed in section \ref{scutoff}
we speculate that such a slight breaking of locality might give us the hope
of obtaining anultravioletly cut off theory in spite of the these manifold or
projective space-times are a priori having infinitely small distances on the
same footing as the large ones.

In secton \ref{sHuge} we put forward the remark that our model
of the imbedding into a projective space time might give some hope for
being able to explain the appearance of the huge almost flat space time volumes,
we find phenomenologically; and in section \ref{sChar} we deliver the
speculation of characterising a projective space as a set/space on which a
froup acts in an especially strong way. To say sharply $n$-transitively
is connected with problems in as far as there are no true more than 
3-transitive, but anyway...
infinite spaces.

In section \ref{sC} we conclude and resume the article.

  \section{Astri Kleppes and mine Theorem:}
  \label{theorem}

  Even taking an action $S[\phi]$ depending on many fields defined over a
  space-time manifold {\bf not to be a priori local at all} but only to obey
  \begin{itemize}
  \item{1.} $S[\phi]$ is Taylor-expandable as a {\bf functional},
  \item{2.} It is {\bf symmetric} under the {\bf diffeomorphism symmetry},
  \item{3.} We observe it only with so long wave lengths that only products
    of fields up to some limited dimension is observed,
  \end{itemize}
  then the effectively observed theory will have a weak form of locallity, in
  the sense, that the action will be observed as one of the form
  \begin{eqnarray}
    S[\phi] &=& F(\int {\cal L}_1(\phi(x))d^dx, ...,
    \int {\cal L}_n(\phi(x))d^dx), 
    \end{eqnarray}
  meaning the action functional $S[\phi]$ would be a {\bf function} of
  a series of usual local action integrals $\int {\cal L}_i(\phi(x))d^dx$,
  but presumably not itself of this form.


  {\bf The field $\phi(x)$ is a common name and can stand for fields with
    many different transformation properties under the diffeomorphism symmetry}

  The field $\phi(x)$ is just short hand for any of the many fields we know
  (or do not even know) like
  \begin{eqnarray}
    \phi(x) &=& g^{\mu\nu}(x), V_a^{\mu}(x), \varphi(x), A_{\mu}(x), \sqrt{g},  ...
  \end{eqnarray}
  or even combinations(products) of them 
  with their various transformation properties under diffeomorphism symmetry.

  \vspace{1mm}

  With diffeomorphisms we think of bijective maps of the manifold on which the
  theory is defined onto itself having the property of being many times
  continuous differentiable.

  They are smooth deformations of the manifold $M$ say.

  {\bf A Diffeomorphism transforms bijectively the Manifold and is continuos
    differentiable (some number of times)}
  
  \includegraphics[scale=0.7]{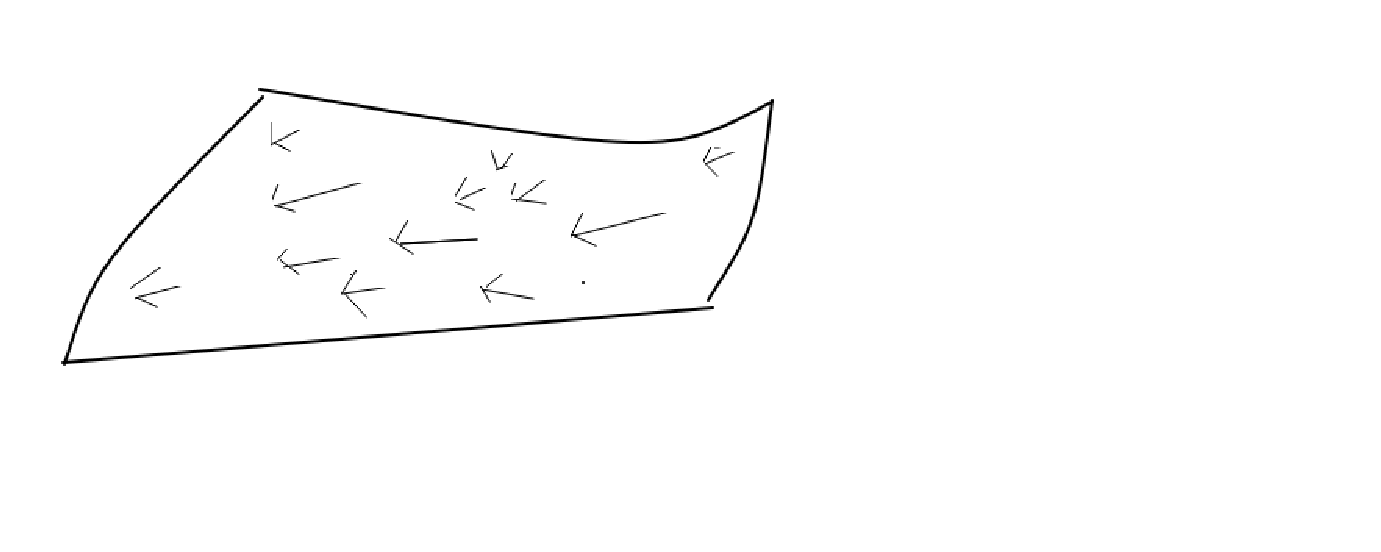}

  {\bf We can likely (almost) do with less than full diffeomorphism
    symmetry}

  We believe the argument for locallity (which I still owe you) could go
  through with similar symmetry such as:
  \begin{itemize}
  \item A projective space time with as symmetry the projective maps
    of this space onto itself.( We shall give the reader a reminder about
    projective spaces in section \ref{sprojective}.)
  \item The symplectimorphisms of a non-commutative space-time with a
    symplectic structure on it.(The symplectomorphisms are maps
    preserving the/a symplectic structure defined on the space.
    \cite{symp})
  \item Of course a true manifold with its diffeomorphisms is o.k.
  \end{itemize}

  But a Minkowski space time {\bf with a distance} between two points that
  cannot
  be varied by the symmetry of the goemetry would {\bf not} be suitable for
  our
  derivation of locality.

  {\bf Essential is that you by the symmetry can move one point around even
    keeping another point fixed}, so that the {\bf only} kept
  information on the relative position of a pair of points is if they coincide
  or not.

  {\bf We  did NOT get full/true Locality out: Only $S[\phi]$ of the Form
    $F(\int {\cal L}_1(x)d^dx, ...,\int {\cal L}_n(x)d^dx)$.}

  With Random Dynamics derivations you are often {\bf not quite} successful
  as here:

  \begin{itemize}
  \item {\bf True locality:} $S[\phi]= \int {\cal L}(x)d^dx$.
  \item {\bf But Only got:} $S[\phi] = F(\int {\cal L}_1(x)dx,...,
    \int {\cal L}_n(x)dx )$
  \end{itemize}

  But that is precisely {\bf interesting} because then the suggestion is
  that nature may only have the {\bf not quite} successful form of
  e.g. locality:

  In fact it is suggested: {\bf The coupling constants such as the fine
    structure constant or the Higgs mass or the cosmological constant  ``knows''
    about what goes on far away.}

  {\bf The Coupling Constants ``knowing'' about Remote Happenings is an
    Advantage}

  That the effective coupling constants or the cosmological constant etc.
  depend on integrals like $\int {\cal L}_i(x)d^dx$ gives at least
  {\bf hope for solving finetuning problems} such as:

  Why is the cosmological constant so phantastically small - in say
  Planck units -?

  Now we can at least hope it is so small to make universe big or
  flat...{\bf But if the coupling constant (say cosmological constant) did
    not ``know'' about the remote, it could not adjust to it.}

  Another fine tuning problem is: Why is the Higgs mass and thereby the weak
  scale so small compared to Planck scale or unification scale (if there
  were unification)?

\section{The Propagation Problem, Need for Gravity}
\label{propagation}

{\bf Another Trouble for it being a full Success: No Propagation
  without a spontaneous
    breaking $g^{\mu\nu}$. I.e. $<g^{\mu\nu}> \ne 0$}

  If there were no spontaneous breakdown, so that in vacuum all fields $\phi$
  had zero expectation value, then there would be {\bf too much locality},
  superlocality:

  There would not be place for useful derivatives in the Lagrangian density
  ${\cal L}(x)$, because $\partial_{\mu}$ could only be contracted to the
  $dx^{\mu}\Lambda dx^{\mu}\Lambda \cdots dx^{\rho}$, but that gives only an
  action which is a boundary integral only (integral of total derivative).

  {\bf So a gravity field with vacuum expectation value is needed}
  (smells like deriving gravity as needed at least).

  {\bf Viewing Gravity from reparametrisation Invariant Fundamental Theory}

  {\huge Gravity or Physics with propagation, needs a breaking of scale
    invarince } since the equation of motion
  \begin{eqnarray}
    g^{\mu\nu}\partial_{\mu}\partial_{\nu}\phi&=& 0 
  \end{eqnarray}
  needs an upper index $g^{\mu\nu}$ for being reparametrization invariant.

  A non-zero $g^{\mu\nu}$ represent a spontaneous break down of a symmetry
  involving say scaling or reparametrizations. Similar ideas by \cite{Percacci}.


  {\bf The theorem: Spontaneous breaking of Reparametrization Symmetry Needed}

  {\bf Theorem:} {\em If a theory with reparametrization invariance is
    not spontaneously broken - meaning the vacuum is totally
    reparametrization invariant - then propagation in this world is impossible.}

  The speculative suggestion: If {\bf reparametrization} transformations
  constitute
  a {\bf fundamental symmetry}, there would be no waves going from one point
  to another;
  so only to the extend that vacuum has some {\bf breaking} of the symmetry of
  {\bf reparametrization} we can get propagation. So not much interesting
  physics
  could go on without this spontaneous breaking. {\bf Gravity-like fields}
  - basically $g^{\mu\nu}$ - non-zero in vacuum are {\bf needed}.   

 \subsection{ New Way of Reducing Dimensionality, use Degenerate $g^{\mu\nu}$}
  \label{dreduction}
  
  {\bf Side remark on Dimensional Reduction in the Spontaneous Breakdown}

  Even if dimension $d$ of the fundamental space were high, we could
  have that the rank of the $g^{\mu\nu}$ tensorfield in vacuum was lower.
  In that case the world in which we could propagate would be of the lower
  dimension.

  But say spinor-fields  would - yuo might think - anyway have to have
  numbers of components
  matching
  the fundamental dimension, but alas: There is no spinor representation
  of the general linear group which the transformation group of the tangent
  space for the symmetries, we care for in this talk! So can a model
  with reparametriztion invariance as fundamental symmetry have spinors at all?

  Well, even in ordinary generel relativity, we know that the spinor fields
  are indeed w.r.t. curved space indices scalars - they only have the
  so caled flat indices, which are really only eneumerations of various
  vierbein fields - so they are indeed {\bf scalars}. This means that
  without the vierbeins the fermion fields could not propagate. So realistically
  we have to think of the breaking of the too large symmetry comes by
  means of vierbein fields rather than by a true $g^{\mu\nu}(x)$ alone.
  In fact the most likely model is presumably that there is bound state
  combination of the vierbeins making up the $g^{\mu\nu}(x)$.

  \includegraphics[scale=0.5]{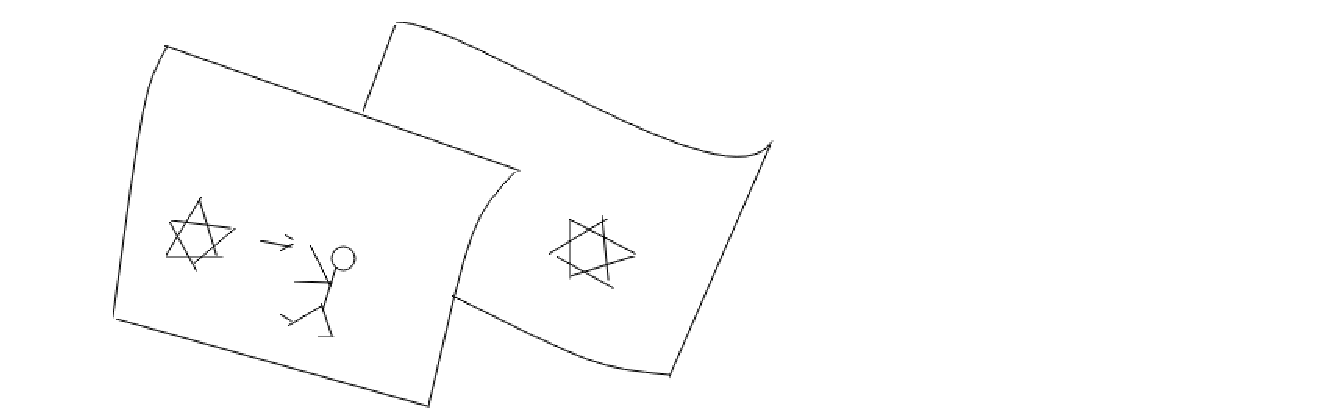}

  {\em On the figure is seen a couple of layers imbedded in the manifold;
    a couple of layers with no communication possible between them,
    in the case of
    degenerate $g^{\mu\nu}$, i.e determinant =0}  
  
  If in vacuum the rank of the upper-index metric tensor/matrix is lower
  than the dimensionality of the manifold, then there appears surfaces on which
  signals can propagate, but from surface to surface it cannot.

  {\bf We imbed Gravity and the rest into Spacetime WITHOUT metric, as Just
    Manifold or Projective Space or some Noncommutative phase space, ...}
 
The idea of imbedding the gravitational manifold from general relativity into
an imbedding space is an old one, see e.g. \cite{Paston}. But Sheikin and
Paston imbed the general relativity space into a flat metric space time. In
the present work we are interested in {\bf imbedding into a geometrical space},
which have {\bf no metric} but rather has such symmetries that locally
it is part
of the symmetry that a small neighborhood can be deformed and scaled up or
down in size, so that a metric would be forbidden by the symmetry and
at best be allowed as spontaneous symmetry breaking. The space in which to
imbed in our present work  is rather thus thought upon as either a {\bf pure
manifold} with no further structure {\bf or  a projective space-time}.

  {\bf In spaces with local linear deformation like: Reparametrization
    invariant or Projective space, No Signature}

  It is the $g^{\mu\nu}(x)$ that has the signature - in physical world
  3+1 -, so without $g^{\mu\nu}$ Minkowski and Euklidean spaces are the same.

  {\bf Our Point of View: Start with a locally  Scaling and deformation
    Containing Symmetry.}

  Having in mind our work with Astri Kleppe\cite{locality, locality2} of
  {\bf deriving locality of the
  theory from a reparametrisation} invariant theory - though allowed not to be
  invariant under variations of the measure - for an extremely general action,
  we suggest to assume a symmetry involving - at least locally - such a
  reparametrisation invariance to be assumed before e.g. getting gravity.

  That is to say: We want to assume either reparametrisation invariance or
  something like that, and {\bf after that} hopefully derive or understand
  gravity
  and locality.

  {\bf Geometries with (local) scale and deformation symmetry.}

  Examples of how you can have local deformation symmetry:

  \begin{itemize}
  \item {\bf Full Reparametrization Symmetry}
    This is the symmetry of coordinate shifts in the General relativity.
  \item{\bf Projective geometry space}
    The symmetry of the projective space is a smaller group than that
    of general relativity. (I am personally especially attached to projective
    geometry, because I made my living from teaching it for 6 years.).
    M. Ninomiya and me\cite{projectiveline}.
    \item {\bf Symplectomorphic invariant space.}
  \end{itemize}

  By local scale symmetry we mean that there are symmetries, so that in the
  tangent space to any point we have symmetry under scaling up by any (real)
  factor this tangent space.

  In the present work we really want to have {\bf locally not only scale
    invariance
  but invariance under any real linear transformation $GL(d,{\bf R})$}.

  {\bf Starting Point: Locally General Linear Transformation Symmetry}

  Our starting assumption - in this work - is that there are such symmetries
  assumed that for every point on the manifold you must have symmetry under
  general linear transformations in the tangent space:

  \begin{eqnarray}
    (dx^1, dx^2, ...,dx^d) &\rightarrow & ((Adx)^1, (Adx)^2, ..., (Adx)^d)\\
    &=& (A^1_{\mu}dx^{\mu},A^2_{\mu}dx^{\mu},..., A^d_{\mu}dx^{\mu})\\
    \hbox{for any real matrix} A^{\mu}_{\nu}&\in& {\bf M_{d\; \times\; d}}
    \hbox{(in the curled indices)}.
  \end{eqnarray}

  Having such a symmetry will be enough for guaranteeing that we for a general
  functionally analytic action $S[fields]$ shall get
  locallity\cite{locality, locality2}.

\subsection{Resume propagation}
\label{resume}

  {\bf Propagation Requires Breaking of the Locally General Linear Symmetry}
\vspace{-1mm}
  Usually the propagation of particles in say free approximation is given by
  a D`alembertian equation of motion
  \begin{eqnarray}
    (\Box + m^2)\phi &=& 0\\
    \hbox{but to have local general linear transformation}&&\hbox{invariance: }
    \nonumber\\
    ( g^{\mu\nu}\partial_{\mu}\partial_{\nu}+ m^2)\phi&=& 0\\
    \hbox{we need the $g^{\mu\nu}$ !}&&\nonumber
  \end{eqnarray}

  If such a $g^{\mu\nu}$ is non-zero in vacuum, we have a {\bf spontaneous
    break down} of the symmetry, because the $g^{\mu\nu}$ field transform
  non-trivially under the local general linear tranformations.

  {\bf So we only can propagate (normally) any particles, provided we break
    (spontaneously) this locally general linear symmetry!}

\section{Locality Derivation Argument}
\label{Locderivation}

  {\bf Reviewing our `` Derivation of Locality''}

  Let us review Astri Kleppes and HBN's \cite{locality, locality2}
  ``derivation'' of locality under the assumptions of {\bf diffeomorphism
    symmetry} (invariance under reparamterizations) for a very general
  action $S$  not being a priori local but rather only having the
  diffeomorphism
  symmetry and being {\bf Taylor expandable } in {\bf local fields}:
  \begin{eqnarray}
    S[\phi]&=& \sum_n \frac{1}{n!} \sum \hbox{(or integral)}
    \frac{\partial}{\partial \phi(x_1)} \cdots
    \frac{\partial}{\partial \phi(x_n)} S[
      0] \phi(x_1)\cdots \phi(x_n),\nonumber 
  \end{eqnarray}
  where here $\phi$ stands for  very general fields, possibly with many
  indices, and $S$ {\bf is an a priori non-local action.}

  (``Mild assumptions'': Taylorexapndability, Finite order Lagrange term only
  observed (low energy))

  {\bf Setting for the Derivation of Locality of the Action $S[\phi]$}

  The field $\phi(x)$ can be so general, that it can stand for all the
  fields, we know, or do not know yet
  \begin{eqnarray}
    \phi(x) &=& A_{\mu}(x), g^{\mu\nu}(x), \psi_{\alpha}(x), ... 
  \end{eqnarray}
  $x$ is a coordinate point, but in the spaces like manifold, or projective
  space, there is always a coordinate choice needed.

  The action $S[\phi]$ is a functional of the fields $\phi$ and is assumed
  \begin{itemize}
  \item Taylor expandable (functional Taylor expansion)
  \item But {\bf not assumed local}, since it is the point to
    derive/prove {\bf locality}
    \end{itemize}

  {\bf The Taylor Expansion for Functional in Integral form}

  The {\bf functional Taylor expansion} in the more functional notation
  (i.e. without imagining a lattice cut off say):
  \begin{eqnarray}
    S[\phi]&=& \sum_n \frac{1}{n!} \int
    \frac{\delta}{\delta \phi(x_1)} \cdots
    \frac{\delta}{\delta \phi(x_n)} S[
      0]_{\mu\cdots \nu} \phi(x_1)\cdots \phi(x_n)
    dx_1^{\mu}\cdots dx_n^{\mu}.\nonumber 
  \end{eqnarray}
  Here the $\frac{\delta}{\delta \phi(x)}$ means functional derivative 
 
  {\bf The Crucial Point: All Points can by Symmetry (Difeomorphism symmetry) be
    brought into Any Other one, Trasitivity}

  When there is no distance a priori in the just manifold with diffeomorphism
  symmetry or the projective space, you at least, if you do not go to higher
  order interaction with several fields multiplied, {\bf field at one point
    will interact the same way with fields at any other point, except the
    very point itself.} Thus you either interactions between all points,
  or interaction of the fields at the same point, i.e. locality.

  {\bf Spelling a bit out the Functional Taylor Expansion}
  
 The {\bf functional Taylor expansion} in the more functional notation:
  \begin{eqnarray}
    S[\phi]&=& \sum_n \frac{1}{n!} \int
    \frac{\delta}{\delta \phi(x_1)} \cdots
    \frac{\delta}{\delta \phi(x_n)} S[
      0]_{\mu\cdots \nu} \phi(x_1)\cdots \phi(x_n)
    dx_1^{\mu}\cdots
    dx_n^{\mu}.\nonumber 
  \end{eqnarray}
The symbol $\delta/\delta \phi(x)$ means functional derivative.

  \begin{itemize}
  \item If we did not allow $\delta$-functions so that we could get no
    contribution  from cases where two of space-variables $x_i$ and $x_j$ say
    coincide, and if there were a symmetry like a translaton $x_i\rightarrow
    x_i +a$ for all vectors $a$ the terms in the functional Taylor expansion
    could only depend on $\phi(x_j)$ via the intergal $\int \phi(x_j) dx_j$.
    So the whole taylor expanded action would be a function of
    such integrals $\int\phi(x)dx$.
    (for the type of symmetries we want to assume one could not even construct
    such integrals, except if $\phi(x)$ transform as a density- like
    $\sqrt{g}$)
    \end{itemize}

   In the  {\bf functional Taylor expansion}:
  \begin{eqnarray}
    S[\phi]&=& \sum_n \frac{1}{n!} \int
    \frac{\delta}{\delta \phi(x_1)} \cdots
    \frac{\delta}{\delta \phi(x_n)} S[
      0]_{\mu\cdots \nu} \phi(x_1)\cdots \phi(x_n)
    dx_1^{\mu}\cdots
    dx_n^{\mu}.\nonumber 
  \end{eqnarray}
  we can {\bf without delta functions $\delta(..)$ } only get the Taylor
  expanded (action) $S[\phi]$ to depend on integrals of the type
  \begin{eqnarray}
    \int \phi(x) dx^{\mu}&&,\hbox{which are invariant under the symmetries}\\
     &&\hbox{(of a manifold say)}
  \end{eqnarray}
  \begin{itemize}
  \item But if we allow $\delta$-functions - in the functional derivatives
    of the functional to be exapnded $S[\phi]$ then one can get integrals
    allowed involving products of fields $\phi(x^{\mu})$ taken at the {\bf same}
    point.
    \end{itemize}

  {\bf What can occur, if we allow the $\delta$-functions ?}

  If we allow the $\delta(...) $ functions and require the symmetry of the
  diffeomorphism of the manifold, the integrals on which the being
  Taylor-expanded quantity $S[\phi]$ can depend, must be integrals
  {\bf symmetric under the
    prescribed (diffeomorphism) symmetry} of {\bf the form}
    \begin{eqnarray}
    S_i[\phi] &=& \int {\bf L}_{i \; \mu\nu\rho\kappa}(x)dx^{\mu}\Lambda dx^{\nu}
      \Lambda \cdots
      \Lambda dx^{\kappa}
    \end{eqnarray}
    where now the ${\bf L}_i(\phi(x), ...,\phi(x))_{\mu\nu\rho\sigma}$ is a product
    of several of the $\phi(x)$ fields at the {\bf same} point $x=x^{\mu}$. If
    these
    arguments for the $\phi(x)$ are not at the same point, then you can by
    the diffeomorphism symmetry transform them around separately and the
    integral would be required to {\bf factorize} into integrals each only
    having fields from {\bf one point} (kind of  locallity).

  \includegraphics[scale=0.9]{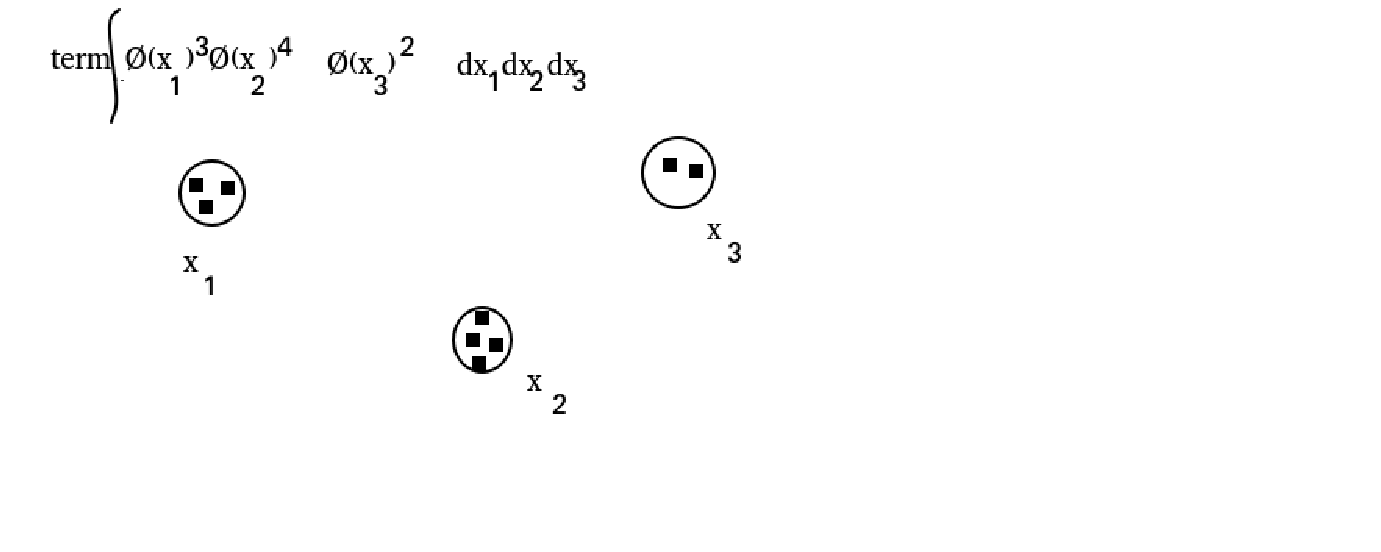}

  {\bf The  $S_i[\phi] = \int {\bf L}_{i \; \mu\nu\rho\kappa}(x)
    dx^{\mu}\Lambda dx^{\nu}
      \Lambda \cdots
      \Lambda dx^{\kappa}$ are ordinary local Actions, but...}

  We did {\bf not} derive that the action functional we discussed $S[\phi]$
  was of the form  $S_i[\phi] = \int {\cal L}_{i \; \mu\nu\rho\kappa}(x)
  dx^{\mu}\Lambda dx^{\nu}
      \Lambda \cdots
      \Lambda dx^{\kappa}$, but {\bf only that it could only depend on the
        fields via such integrals}. So rather only the expanded action
      is a {\bf function} of such integrals.


  {\bf We derived a form of the Diffeomorphism Invariant Taylor Expandable
    Action $S[\phi]$ as a Function $F(S_1[\phi], S_2[\phi],...,S_n[\phi])$
    of usual {\em local} action integral.}

  Indeed the terms in the functional Taylor expansion will be of the form that
  groups of factors are at the same point inside the groups, and that then
  these point are integrated over all the space(manifold). Denoting the
  possible integral over local field combinations
  \begin{eqnarray}
    S_i[\phi] &=& \int {\cal L}_{i \; \mu\nu\rho\kappa}(x)dx^{\mu}\Lambda dx^{\nu}
      \Lambda \cdots
      \Lambda dx^{\kappa}
  \end{eqnarray}
  we get the {\bf form}
  \begin{eqnarray}
    S[\phi]&=& F(S_1[\phi],...,S_n[\phi]).
  \end{eqnarray}
  This form was studied by my student Stillits\cite{Stillits}

  {\bf But this was not Really Locality!}

  The form, which we derived from the diffeomorphism invariance
  \begin{eqnarray}
    S[\phi] &=& F(S_1[\phi], ..., S_n[\phi])\label{unwanted}
  \end{eqnarray}
      {\bf is not truly local; we should have had a linear combination }
      \begin{eqnarray}
        S[\phi] &\stackrel{=}{wanted}& a_1S_1[\phi] + ...+ a_nS_n[\phi].
        \label{wanted}
      \end{eqnarray}
      However, if we construct the {\bf equation of motion} by putting the
      functional
      derivative of (\ref{unwanted}) to zero, we get the wanted
      form (\ref{wanted}).

  {\bf Equations of Motion got Already Local, but...}

  The equations of motion for an action of the form - derived from the
  diffeoemorphism symmetry
   \begin{eqnarray}
    S[\phi] &=& F(S_1[\phi], ..., S_n[\phi])\label{unwanted}
   \end{eqnarray}
   becomes
   \begin{eqnarray}
     0&=& \frac{\delta S[\phi]}{\delta \phi(x)}\\
     &=& \sum_i \frac{dF}{dS_i}|_{\phi}*\frac{\delta S_i[\phi]}{\delta \phi}\\
     &=& \frac{\delta}{\delta \phi}\sum_i  \frac{dF}{dS_i}|_{\phi}* S_i[\phi]
       \hbox{ considering $\frac{dF}{dS_i}|_{\phi}=a_i$ constants}\nonumber
     \end{eqnarray}

  {\bf The coefficients $\frac{dF}{dS_i}|_{\phi}$ do depend on the fields
    $\phi$, but integrated over all time and all space.}

  Effectively these coefficients
  \begin{eqnarray}
    a_i &=& \frac{dF}{dS_i}|_{\phi}
    \end{eqnarray}
  to the various possible local actions $S_i[\phi]$ {\bf do depend on the
    fields $\phi$} but since they depend via  integrals over all space time,
  we can in pracsis take them as constants. Indeed they are the {\bf coupling
    constants} which we just fit to experiments. But it means that our lack
  of completing the derivation of locallity means:
  {\bf The coupling constants - say fine structure constant etc. - depends
    huge integrals over space time}, although composed in a way which depends
  on the fundamental non-local action, which we do not know (yet?).

  {\bf We did not get full locallity! Coupling constants depend on all
    space-time }

  We got, that {\bf the Lagrangian density only depends on the fields
    in the point you write this Lagrangean density}, and that is practical
  locality, {\bf but we did not get locallity for the coupling constants in
    the sense that they with our derivation ``know about'' what goes on all
    over space and time, including even future.}

  .
  This suggests that the question of what the {\bf initial conditions} should
  be at least in principle needs an extra discussion.

  In principle there is a backreation for any choice of initial condition
  because it influences the couplings depeding on the development of it
  also to far future.

\section{MPP}
\label{sMPP}

  {\bf Prediction of Severel Degenerate Vacua (``Mulitple Point
    Criticallity Principle'')}

  This great importance of integrals over all space time of the fields
  could very easily lead to limitations for such overall space time integrals.

  Such specification of a non-local (even in time) is analogous to
  {\bf extensive quantities} in thermodynamics:

  If you specify them you risk to put your system into a phase transition point.

  If you specify several of them you easily end up with several phases in a
  ballance.

  Here the analogues of the {\bf intensive quantities} like temperature
  and chemical potentials are the coupling constants, which with our
  incomplete localliy derivation depend on what goes on or will go on or
  has gone on in the universe.

  {\bf When Ice and Water in equilibrium Temperature $0^0$}

  When one has the situation of slush - that there is both water and ice -
  then ones knows the temperature must be zero, cold but not extremely cold.
  
  \includegraphics{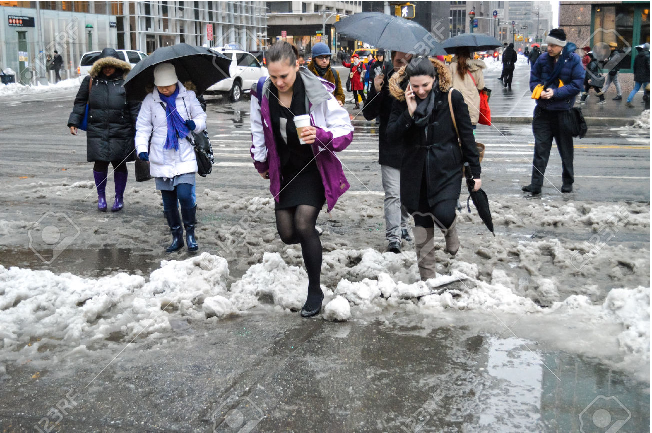}

  In the analogous way we have what we called {\bf the Multiple Point
    Criticalallity Principle} when in space time one has several vacua in
  ballance taken to mean, that they have the same energy density.
  To find a good argument for this suggested principle we speculated that
  the some integrals of the type $S_i[\phi]$ got specified values analogously
  to fixing extensive quamtities in chemistry. Then it could easily be that
  the specified quantities could only be realized when there were indeed
  several (vauum) phases anlogous to the some specified combinations of a
  number of mols water, total energy and volume could enforce there to
  be sluch and may be even water wapor and the temperature and pressure
  could be enforced to be at the triple point.

  With the Action derived from our locality derivation involving strongly
  the many integrals over all space time, one could easily imagine that
  by some way of getting a selfconsistent solution it could turn out that
  several of these integrals get so restricted, that one has such a situation
  similar to the slush one, that it was needed to have several phases of
  vacuum in space-time.And if they should be in equilibrium by having
  the same energy densities say, then coupling constants might end up in some
  critical point where the phases could coexist.

  At least we can say, that since such integrals, as $S_i[\phi]$, appear in the
  form we argued
  for, seeking consitent solutions for the equations of motions
  could easily lead to restrictions. In fact to get consistent
  solution to not quite local equtions of motion is not at all trivial.
  The timedevelopment, namely, to vaues of the inegrals over space and time,
  thus giving vaues for the effective couplings.

  So it is not unlikely that our not quite local action would lead to
  our earlier proposed {\bf multiple point criticallity principle.}
  
  This  would be a success, if we could get the Multiple point criticallity
  principle out as extra premium from the attampt to derive locallity, because
  Colin Froggatt and claim to have PREdicted the mass of the Higgs boson before
  the Higgs boson were found experimentally, by means of the multiple
  point criticallity principle. In fact in the article Phys.Lett. B368 (1996)
  96-102
  following an arXiv-article submitted in Nov. 1995, we publiched the Higgs-mass
  prediction $135 GeV \pm 9 GeV$ also seen on the following painting:  
  
  {\bf We PREdicted the Higgs Mass by Several Balancing Vacua (MPP) before
    the Higgs was found}
  
  \includegraphics[scale=0.32]{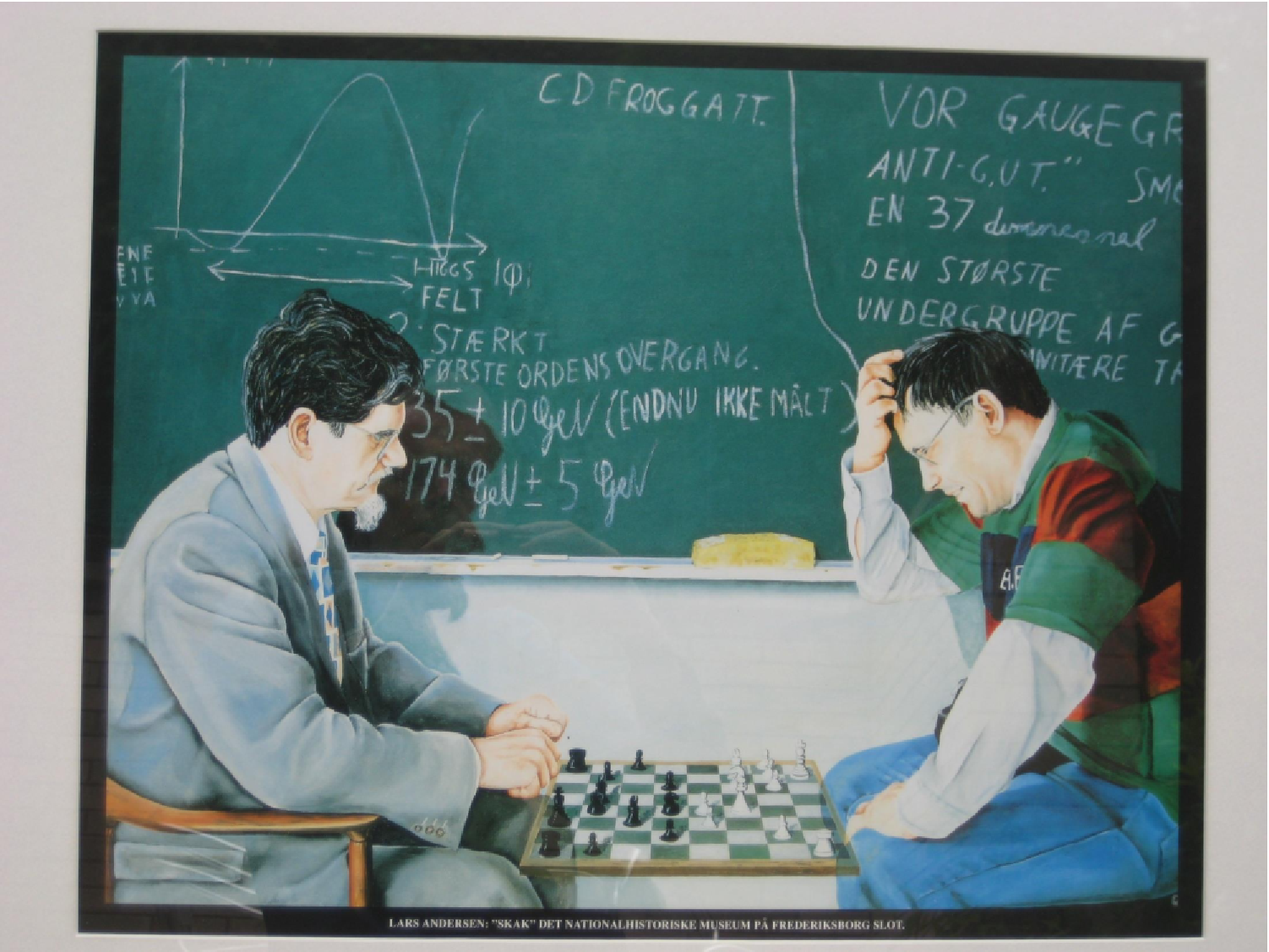}

  The painting of me together with the Danish finance minister - whom I
  only met many years later - were painted in the 90's much
  before the
  Higgs was observed in LHC (=Large Hadron Collider) with 3 $\sigma$ in 2012
  and finally estblished in 2014. Neverthe less you can see the mass of the
  Higgs particle written as $135 GeV$ $\pm$ $10 GeV$ (only the 1 is hidden
  behind Mogens Lukketofts head), but in our article in Phys. Rev. we have the
  $135\pm 10$ GeV. (The measured mass turned out 125 GeV).

  This we like to take as a support for the multiple point criticallity
  principle, and thus if this could be a consequence of the incompletely
  local action form for that even this form is being a little supported.

  Actually it would be rather impossible to see how such phases with same
  energy density could come about in a world with complete locallity.

  If namely one vacuum did not appear before after some time in the universe
  development - and that must be so because there were so hot in the beginning,
  that there were no vacuum proper anywhere - then how could any coupling
  constant or the Higgs mass adjust to make such a vacuum obtain a special
  value for its energy density say, when the vacuum had yet never existed ?
  At least it looks that some ``non-locality'' of this type must exist:
  Higgs mass or other parameters in the theory such as coupling constants
  and the cosmological constant must have been informed from the beginning
  about e.g. vavuum properties of vacua first existing long after.

  This type of lacking locality is precisely the one we did not mannage to
  derive.
  
\section{Trasitivity}
\label{Trans}

  {\bf More on Transitivity}

  When a group $G$ acts on a space $X$
  \begin{eqnarray}
    \alpha : G\times X &\rightarrow& X\\
    \hbox{denoting } \alpha(g,x) &=& gx\\
    \hbox{so if }gx_1 &=& x_2,
  \end{eqnarray}
  it means the group element $g$ brings the element $x_1\in X$ into $x_2\in X$,
  then we say {\bf $G$ acts $n$-transitively provided there for any
    two ordered sets of $n$ different points in $X$, $(x_1,x_2,...,x_n)$ and
    $(y_1,y_2,...,y_n)$ exist a group element $g\in G$ such that
    \begin{eqnarray}
      gx_i &=& y_i \hbox{ for all $i$.}
    \end{eqnarray}}

    We say it is sharply $n$-transitive, when this $g$ is unique.

  {\bf $d$-dimensional projective space has a symmetry group acting almost(!)
    $(d+2)$-transitively}

  Examples:
  \begin{itemize}
  \item Under the action of diffeomorphisms on a manifold the action is
    n-transitive for any integer n; but is far  from being sharply
    n-transitive. (There is indeed a theorem  generalized by J. Tits and
    M. Hall, who proved that there are
no infinite sharply n-transitive groups for n$\ge$ 4.See e.g.\cite{Simonov})
  \item In $d$-dimensional  projective space $PS(d,{\bf R})$ the symetry
    group acts essentially $(d+2)$-transitively, but not truly so, because
    the immage of points say on a line remains on a line.
    Only the projective line $PS(1,{\bf R})$ is truly 3-transitive.
  \item Euclidean spaces are only (1-)transitive under their symmetry.
    It is the translation group that acts transitively. When the group
    conserves the length of line there can be no even 2-transitivity.
    \end{itemize}

\section{Repeat}
\label{sRepeat}

  {\bf Repeating Argument using Action}

  Instead of looking at the equation of motion we could ask, if we could
  make an action
  \begin{eqnarray}
    S[fields] &=& \int {\cal L}_{\mu\nu\rho\sigma}(x)dx^{\mu}\Lambda dx^{\nu}\Lambda
    dx^{\rho}\Lambda dx^{\sigma},
  \end{eqnarray}
  which is invariant under our symmetry having locally generallinear symmetry
  and at the same time can describe a propagation of some fields.

  If a field $\phi$ shall not be determined locally by the other fields, but
  appear in equation(s) with derivatives, there must be a derivative acting
  on $\phi$ i.e. say $\partial_{\mu}\phi$ occuring in the Lagrangian
  density ${\cal L}_{\mu\nu\rho\sigma}$; but with what to contract the
  lower index $\mu$ on $\partial_{\mu}\phi$ ? To some field with an upper
  curled index like a vierbein $V^{\mu}_a $ or a $g^{\mu\nu}$? Yes but
  if we work in vacuum and there were no spontaneous break down of the
  symmetry these fields would be zero.

  {\bf Continuing repeating Derivation of Need for Spontaneous Breaking of
    Locally General linear symmetry}

  Looking for making 
 \begin{eqnarray}
    S[fields] &=& \int {\cal L}_{\mu\nu\rho\sigma}(x)dx^{\mu}\Lambda dx^{\nu}\Lambda
    dx^{\rho}\Lambda dx^{\sigma},
 \end{eqnarray}
 invariant under the symmetry, but still with fields propagating
 even with vacuum not breaking the symmetry. (We shall show you cannot find
 such an action.)

 Can it help to let the $\partial_{\mu}\phi$ combination be contracted with
 a $dx^{\mu}$ to give it a chance to propagate? 

 In fact
 \begin{eqnarray}
   dx^{\mu}*\frac{\partial \phi(x)}{x^{\mu}} &=& d\phi(x)
   \end{eqnarray}
 is a total derivative. If you now wanted to make the term second order in the
 $\partial_{\mu}$, you would use yet another of the factors in the
 measure
 \begin{eqnarray}
   d^dx &=& dx^1\Lambda dx^2 \Lambda \cdots \lambda dx^{n-1}\lambda dx^n, 
   \end{eqnarray}
 and the second order term would be like
 \begin{eqnarray}
   \frac{\partial \phi}{\partial x^{\mu}} dx^{\mu} \Lambda
   \frac{\partial \phi}{\partial x^{\nu}} dx^{\nu} &=& 0 \hbox{for same
     $\phi$ in the two factors.}\\
   \hbox{ or different $\phi$'s, $\phi_a$ and $\phi_b$ }&&\nonumber\\
   \frac{\partial \phi_a}{\partial x^{\mu}} dx^{\mu} \Lambda
   \frac{\partial \phi_b}{\partial x^{\nu}} dx^{\nu}&=& d\phi_a\Lambda d\phi_b
   \hbox{ a toal derivative.}
   \end{eqnarray}

 Couple it directly to the
 $dx^{\mu}$'s?

 Seems not to give a propagating equation of motion usual type.

  Let us here also remark that to get such an integral over all space time
  as we discuss, one also has to include a quantity that transforms as
  the usually known $\sqrt{g}$ the determinat of the metric tensor with lower
  indices. It is e.g.
  \begin{eqnarray}
    \int \sqrt{g} d^dx&=& ``4-volume''\\
    \int \frac{\partial \phi}{\partial x^{\mu}}*
    \frac{\partial \phi}{\partial x^{\nu}}g^{\mu\nu}\sqrt{g}d^dx&&,  
    \end{eqnarray}
  which are meaningful reparametrization invariant space time integrals,
  while there is no meaningful 4-volume of the manifold nor of the
  projective space time {\bf without the $\sqrt{g}$ or something replacing it.}
    The projective space has no size in itself. Also for that you need
    a spontaneous breakdown.

    In reality is what we need for being able to get a propagation for
    a scalar $\phi$ the combined quantity
    $g{\mu\nu}\sqrt{g}$ where then $g$ is determinant of the
    metric with lower indices being the inverse of the one of the upper
    index.

\section{Projective}
\label{sprojective}

  {\bf In Plane Projective Geometry All (different) Lines Cross in a Point}

  \includegraphics[scale=0.5]{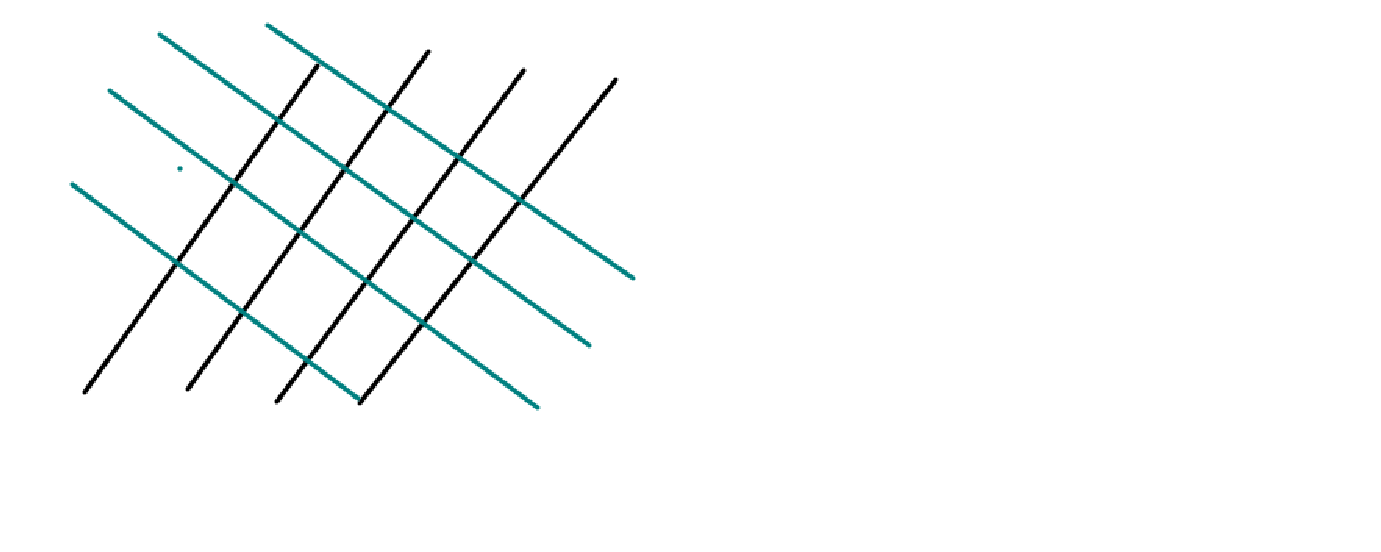}

  {\bf Bundles of parallel lines are identified with points on the line at
    infinity.} {\bf So parallel lines cross there.}

  Really one makes a description of the $d$-dimesnional projective
  space by taking a $d+1$ dimensional vector space and identifuing the
  rays (i.e. sets of vectors proportinal to each other. A class
  of such non-zero to each other proportional vectors is
  called a ray) with the points in the d-dimensional projective space.
  The lines in the projective space are then identified with the
  two dimensional subspaces of the vector space, and the projective
  plans with the thre-dimensional subspaces, and so on.

  It is thus possible to make projective spaces corresponding to different
  fields in as far as onecan have vector spaces with diffeerent field.
  In this article we are interested in using the real field only.
  
\section{Phenonenological Evidence for World being a Projective Space-time}
\label{sPheno}

  {\bf Infinite far out points in opposite direction identified in projective
    geometry}
  
  \includegraphics[scale=0.8]{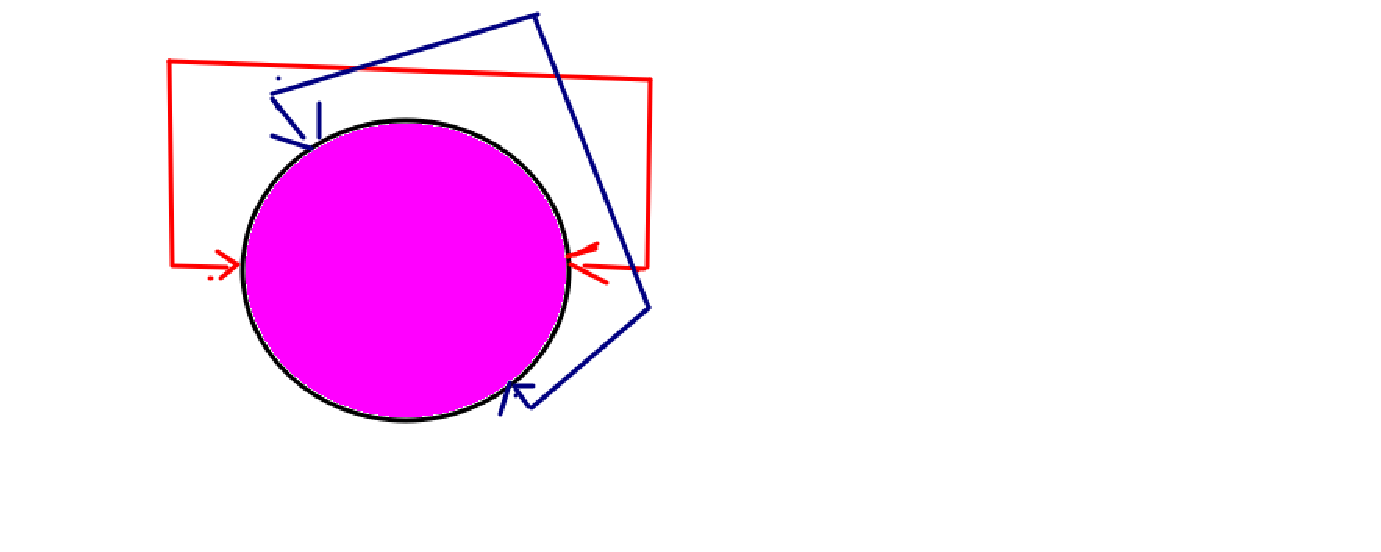}
 
  We here stress, that since in projective geometry a bundle of Parallel lines
  is considered {\bf only one point on the line or plane or etc. at infinity}
  there is no distigtion between the point at infinity in one direction
  along the bundle of parallel lines and the point on the infinite plane or
  watever. There is {\bf only} one point at infinity for each bunch of lines.

  This is illustrated on the fgure by the arrows pointing to identified
  directions so to speak.

  If somehow our universe really were a projective space, then you might see
  the same object in the two opposite directions. That would give of course
  a correlation of e.g. the radiation comming from two opposite almost
  infnities. They would fluctuate in similar way because of being the same
  point on the infinite plane (in three dimensions)

  {\bf Lowest l WMAP fluctuations}
  
  \includegraphics[scale=0.5]{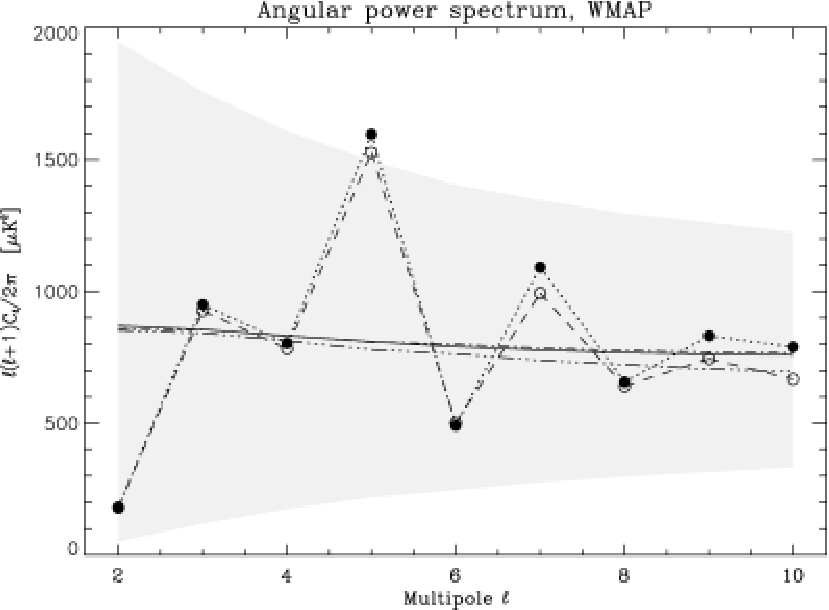}

  The analysis of the microwave background radiation is typically done by
  resolving the fluctuation of the temperature as function of the point
  on the sky into a description expanded on spherical harmonics.
  Thus one presents e.g. the size of such fluctuation connected to the
  various spherical harmonics, which are marked by $(l.m)$. We shall have in
  mind that the even $l$ spherical harmonics have the same value in exact
  opposite directions, while the odd $l$ ones have just opposite values in
  opposite directions.

  On the figure we have the experimentally found fluctuations as function
  of $l$ (averaged over m) for the first few lowest $l$'s.
  
  Remarkable:{\bf Even} l {\bf fluctuatons are relatively low, while the odd}
  l {\bf ones
    are relatively high}

  We know that microwave backgound radiation comes from 13.7 millird light
  years away, so if the universe should really be a projective space, the
  infinite plane or infinite thre-space if we think of the fourdimensional
  space time as projective, sould be not much further away than 13.7 millard
  light years if we should be able toobserve it.

  {\bf If Projective Space ``seen'' in CMB-fluctuations, then Universe Not
    Much bigger than Visible Universe}

  The just shown:
  \begin{itemize}
  \item $Y_{lm}$-proportinal modes in temperature variation over sky with
    {\bf even} $l$ have {\bf lower} fluctuation.
  \item $Y_{lm}$-proportional modes in temperature variation over sky with
    {\bf odd} $l$ have {\bf higher} fluctuation.
  \end{itemize}
  if taken seriously implies that {\bf the visible universe edge is not very
    far from where there is the  identification of the diametrically
      opposite points} (on say the infinite line). So Universe would not be
    so huge as the very accurate flatness would indicate!

    \subsection{Correcting for my mistake}

  {\bf Post Talk Slide: Naively you expect the opposite!}

  I got very confused and chocked, when I persented the foregoing slides,
  because naivly thinking of a three-dimensional Projective space:

  Since the points in opposite directions are {\bf related, in fact the
    same region,} the {\bf even l} spherical harmonics, should show a
  {\bf big} fluctuation, because they add together by continuity across
  the infinite plane related regions, and thus counts really the same
  fluctuation in two ends as statistically independent and thus over estimate
  the fluctuation. Oppositely weighting with an {\bf odd l} spherical harmonic
  you add with a relative minus sign two close by (across the infinite plane)
  region contributions and should get approximately zero. Thus the fluctuations
  for {\bf odd l should be small.}

  {\bf Thus my chock: The ``evidence'' I had believed, had the wrong sign!}

  {\bf We Forgot the Time Direction...}

  We should not have looked for opposite points in the purely 3-dimensional
  space,

  Rather than looking 
  {\bf for the opposite point on the infinite 3-space
we should look for opposite points also in the time direction i.e. in 
    the
    4-dimensional space-time.}

  If we take that what should be Big Bang is rather just  the narrowest point,
in some sort of 
bouncing universe, than we can at least speculate to have a mirror
symmetry in this the narrow universe region.
Let us consider this time usually taken as big bang to be rather a center
/ origo of the space time, in the sense that we consider the opposite ends
of lines extending from this big-bang like region and claim that because
of a three-space at infinity having the  fluctuations are approximately
the same  in the two ends of a
line through this big-bang region .


  {\bf Imagine Imbedding a Bouncing Universe Space-time into a 4-dimensional
    Projective space, filling it out}
  
  \includegraphics[scale=0.7]{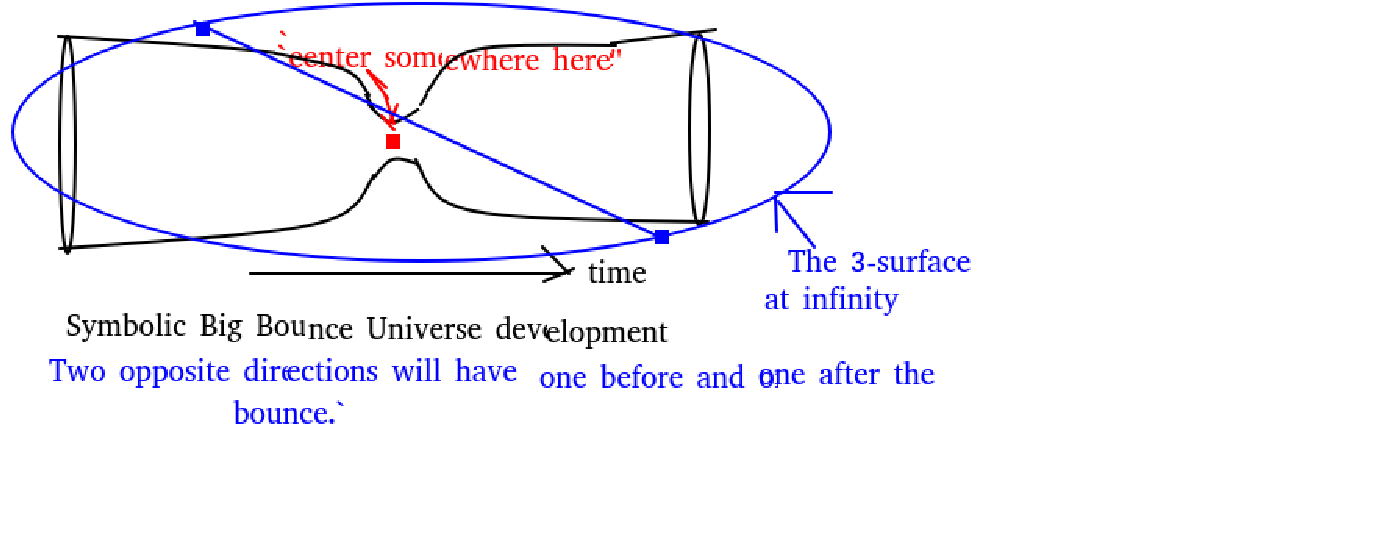}

  This drawing is very symbolic, but of course one can see that the two
  {\bf directions} corresponding to the same
  point on the infinite 3-space, now have opposite times too. That is to say
  that one of them are in the prebigbang time (which probably do not exist,
  but the projective space has no singular start so it goes extremely badly
  with the big bang theory; so to have any chanse with projective space-time
  we better give up big-bang).

  {\bf In Projective Space No Singularities like Big Bang, so better Bounce}

  In the Projective space(s) all points are symmetric with each other
  (trasitivity of the grouop action) and thus no point-singularities, so
  that a {\bf big bang start would not fit well into the projective spacetime}.

  Therefore we rather imagine here a {\bf big bounce model} in which there
  is contracting universe before it reexpands, although such a model
  has rather many problems with second law of termodynamics, and with
  how an about to crunch universe can get its contraction turned to an
  expansion. But an {\bf empty De Sitter space can bounce}, so we might
  postulate
  that in the midle of times the universe is an empty De Sitter space.

  So two opposite time directions in respectively the expanding and the
  chrunching halfs of the sapce time.

  {\bf Inflaton Field goes up to top of Potential to Stand there as Long as
    possible to get preferably slowest roll}

  Let us imagine as our model to get at all a reasonable imbedding into
  the projective space, and a preferably as long as achievable inflation,
  locally the inflaton field in crunch-part of time in the anti-inflaton
  period runs up the potential hill and stops very close to the top of
  the potential. Then it falls down again, first extremely slowly and
  then unavoidably faster.

  {\bf Opposite Point Identified by the 3-space at infinity, are oppsite
    in both space and time.}

  If an event at the recombination era at 370000 years after
  the big bounce is supposed to be sufficiently far out to be close to
  a point on the 3-space at infinity, then this approximating point
  at infinity is identified with a point at infinity in the opposite direction
  in space, meaning on the sky, but it shall also be opposite in time.

  The latter presumably means it shall be on the chrunching time sector
  if we
  have looked at a
  search the point opposite to a point in the expanding time sector.

  {\bf Speculate biggest fluctuation in the Time of reaching the
    Potential Peak}
    
  Let us further assume that the largest fluctuation in the inflaton
  field comes from the {\bf time} at which the inflaton just reaches the peak
  of the potential varies randomly from region to reagion in space.

  If we have points that are relatively far from each other such fluctuation
  of the times of peaking should be strongly fluctuationg.

  Now notice: If it happens early from the point of view of the
  expanding universe then it happens also early by the point of view
  of the chrunching universe, but that has for this compared our expanding
  universe the opposite effect, because of the time reversal.

  {\bf Prediction of the Sign-inversion by the time reflection not detail
    dependent}

  So whether late or early gives larger or smaller CMB radiation, then
  the time reflected point will always give the opposite to the not
  timereversed one. So we will get the odd l speherical harmonics
  get the biggest fluctuation, and the evenones the smallest!

  {\bf Conclusion on CMB-fluctuation Prediction}

  Assuming:
  \begin{itemize}
  \item Bouncing Universe
  \item Time of Inflaton field Reaching the Peak most important
    for the fluctuations in the CMB radiation.
  \item The Chrunching Universe behave Time reversal invariant to the
    expanding one.
  \end{itemize}
  we get:
  {\bf The odd l shperical harmonic modes shal have the largest
    fluctuations}, while the {\bf even l one the smallest,} contrary
  to the intuition forgetting the time to be also reflected and via
  timereveral can give the opposite sign.

  \subsection{Statistics of the CMB deviation for small l}
  We must admit there are only ca. 2 standard deviations from also the
  low $l$ fluctuation observations agreeing  with the staistical model, so
  there is
  only two standard deviations to build the story about the projective space on.
  So it is very weakly supported.
  
  {\bf Only 2 s.d. from statistically understood low l modes}
  In spite of the statistical significance of the observed even-odd
  assymmetry we used to support Projective geometry is only $\sim 2 s.d.$
  theorists sought to explain these low l
  fluctuations, e.g. R. Mayukh et al, \cite{Explaining} by Superstring
  excitations, and by ``Punctuated inflation''\cite{punctuated}.

  \section{Conclusion}
\label{sConclusion}
  We have reviewed an older work by Astri Kleppe and myself deriving the
  locallity form
  \begin{eqnarray}
    S_{eff}[\phi,...] &=& \int {\cal L}(\phi(x),\partial_{\mu}
    \phi(x),...  g^{\mu\nu}(x), ...)d^dx\\
    \hbox{Really at first: }S[\phi] &=& F(S_1(\phi), ..., S_n[\phi]),\\
    \hbox{where }S_i[\phi] &=& \int L_i(\phi(x), ...) d^d x 
  \end{eqnarray}
  for the effective action, {\bf even though the a priori action $S[\phi,..]$
    is not assumed local.} under the ``not mild  assumption'' of
  diffeomorphism symmetry, which though approximatively is suggested to be
  replacable by a projective goemetry or other space which at least locally
  as symmetry can be deformed, spread out or in in different directions.

  Our main interest in the present article was to point out, that in order that
  such a derivation be usefull for phenomelogical physics a gravity-like
  field, say $g^{\mu\nu}(x)$ or better the combination $g^{\mu\nu} \sqrt{g}$ or
  some corresponding vierbeins have to take
  non-zero expectation values in vacuum. Thus gravity gets a status
  as a spontaneous breaking set of fields needed to have at all the rather 
  abstract property of the physics model of there being propagation, or say
  interaction between different points/events even in an indirect way.
  In other words, were it not for the gravity spontaneous breaking, then
  different events in space time would have no interaction with each other.We
  also review that
  the fact, that the derivation is after all not complete, but only leads to an
  effective locality in as far as the effective coupling constants and mass
  parameters in the quantum field theory resulting {\bf depend actually on
    integrals over all space and all time}, so that they are formally not at
  all local. Only when we by a formal kind of swindel take these on all over
  the space time extended integrals depndend ``coupling constants'' as
  constants we obtain the true local theory. (but of course they are constants
  as function of time.)

  This slight lack of full locality we suggested to be usefull for solving
  some finetuning problems, such as the smallness problem of the cosmological
  constant, and it could also lead to the by us since long beloved speculation
  of ``Multiple Point Criticallity Principle'' saying that there are several
  types of vacuum, and that they have the same energy density(= cosmological
  constant).

  More like an outlook for future work we hoped for a specification
  of a manifold, or taken as very similar a projective space, by an
  assumption of
  there being so much symmetry, that the set of geometrical points/events in
  the
  space-time should be transformed under the sharp $n$-transitive action of
  a group $G$ on a system/set  of events $X$. Higher than $n$ equal 3 is
  actually not possible, but one may hope to come over this problem by
  defining a concept of the action of a group $G$ being only {\bf almost}
  $n$-transitive. In fact here are namely classifications that an infinite
  set cannot have a sharp $n$-transitive action for $n$ more than 3.

  Finally we represented an argument, that the small only about 2 standard
  deviations of the microwave background radition fluctuation data from the
  prediction of the Standard Cosmological Model, could be understood as a
  reflection of the imbedding of our space time into a projective space-time.
  It turned out that it being a space-time, and not only the space, is crucial
  for the sign of the effect, the sign of the deviation from the Standard
  Cosmological Model.

  Thus the model, that the space-time is imbedded into presumably a
  4-dimensional
  space-time, which is a projective space, has some tiny phenomenological
  advance for it.

\section{cut off}
\label{scutoff}
\begin{frame}
  {\bf A speculative Attempt to Cut off Gravity}

  Encourraged by the even phenomenological support for an imbedding of
  the space-time in a projective space, and by the derivation of effective
  locallity, even if you do not impose it, we propose to replace the
  {\bf manifold} with its diffeomorphism symmetry, by a projective space, with
  a little less but very similar symmetry, namely of the group of projective
  transformations.

  Then the derivation of {\bf locality will not be as perfect as with the
    diffeomorphism symmetric manifold}, and rather there will be some
  possibility for {\bf non-local terms in the final action being expanded}.

  But this one could consider an advantage, becuase it could function
  as a cut off, having a chance of allowing a theory with less renormalization
  problems.
\end{frame}

  {\bf Dream of Dynamical Lattice Imbedded in Projective Space}

  If some non-local interaction - allowed by the projective symmetry -
  led to a dynamical lattice imbedded in the projective space, we could have
  a dynamical lattice spontaneously breaking the projective symmetry by
  giving it (essentially) a metric.

  It could function as a cut-off (=regularized) gravity, and other fields
  might be restricted to only be able to {\bf propagate} along the lattice,
  because it could be that there would be no truly continuous gravity field
  except the fields on the lattice points.

  It might be not over-optimistic to calculate possibilites for such a lattice
  of points imbedded in the projective space-time.

  {\bf A Lattice proposal Imbedded in a d-dimensional Projective space-time,
    with some special points}

  Since the non-local interaction allowed by the Projective symmetry will be
  by among $d+3$ points, it will be the best chanse to get a suitable lattice,
  having $d+1$ points in the projective space being especially marked.

  We could let these ``marked points'' to be at first the origin and
  $d$ points on the suface at infinity. The lowest number of points for
  which a (non-local) action dependence would be $d+3$, so by fixing as marked
  $d+1$ marked, we would have a favourite lattice relation, when the two
  points further needed for nonlocal action term being allowed could be
  considered neighbors in a more conventional lattice.

  {\bf What determines the Lattice ?}

  We must imagine that the lattice giving the analogue of a vacuum
  must come from minimizing something - the energy or some
  action - and if so, then it would be simplest that each time you have
  the $d+3$ points needed for having a nonlocal interaction, then
  the minmization of whatever puts such $d+3$ points in a certain
  cofiguration having meaning under the projective symmetry.

  So we should only have a few favourite relative positions of such
  $d+3$ points realized again and again in the lattice.

  \includegraphics{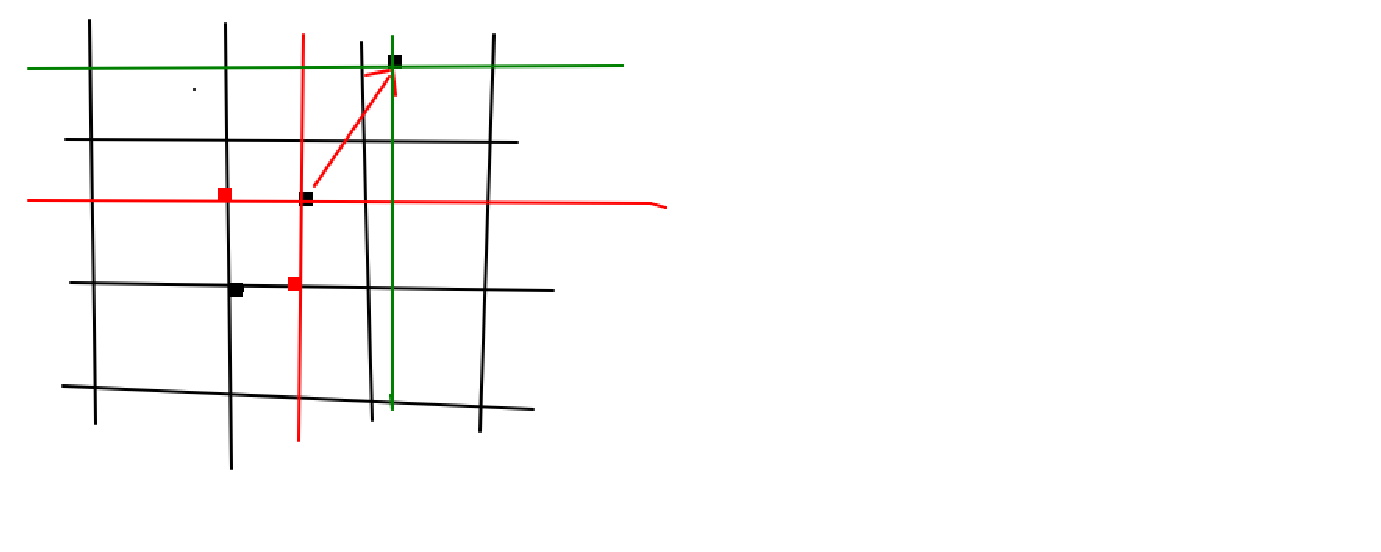}

  {\bf With a Couple (for d=2) of favvourite 5 point combinations we get a
    lattice two dimensional integer coordinat lattice
    (though in log coordinate)}

  It turns out that each step of going from one point to the next by a
  favourite configuration, you step the same step each time in logartithmic
  coordinates.

  But it gives an apriori {\bf flat world} which has a dynamical
  gravity so it can be curved, but first crude prediction is flat.

\section{Huge}
\label{sHuge}
\begin{frame}
  {\bf The Hugeness of the Universe ?}

  Dirac wondered about the huge numbers of order $10^{20}$, that e.g.
  the age of the universe is of the order of $(10^{20})^3$ time the Planck time.

  Assuming a {\bf projective space } background for our space time could
  in an a priori unexpected way enforce the existence of very - infinitely -
  extended space time reagion(s)!

  {\bf Argument goes:}
  
  \begin{itemize}
  \item The projective space of even dimension is non-orientable.
  \item That enforces a hyper-surface, on which the $g^{\mu\nu}$ is
    is of rank one less - say for normal rank 4 it has 3 there.
  \item But then there $g_{\mu\nu}=\infty$.
  \item Appraoching this degeneracy surface the volume relative to the
    coordinates grow so much that an infinite universe in space and time
    pops out.
    \end{itemize}
  \end{frame}
\begin{frame}
  {\bf Non-orientability of Even Dimensional Projective space}

  Most easily seen in the even dimension $d=2$.
  
  \includegraphics[scale=0.6]{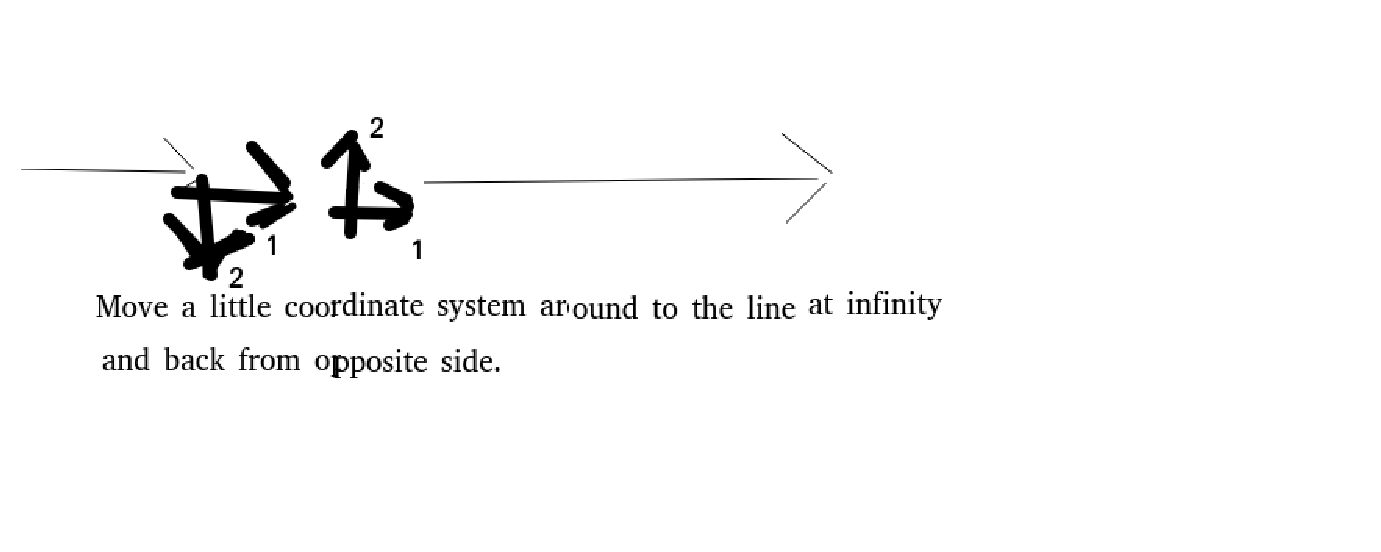}
  \end{frame}
\begin{frame}
  {\bf The Needed $g^{\mu\nu}(x)$ must be degenrate along a 3-surface}

  The determinant $det(g^{\mu\nu})$ cannot avoid a zero surface of dimension 3
  in a 4 dimensional projective space. The sign of this determinant namely
  represents an orientation.

  Write it the coordinates chosen locally $x^1, x^2, x^3, x^4$ and in a certain
  order say 1,2,3,4.
  Then
  \begin{eqnarray}
    \hbox{If } det(g^{\mu\nu}) &>& 0, \hbox{ orientation is that of
      ordered coordinates}\\
    \hbox{If } det(g^{\mu\nu}) &<& 0, \hbox{orientation is 
      opposite
      coordinates in their order}\nonumber
    \end{eqnarray}
\end{frame}
\begin{frame}
  {\bf Think of the determinat $det(g^{\mu\nu})$ followed around to infinite
    line and back the other way}
\includegraphics[scale=0.7]{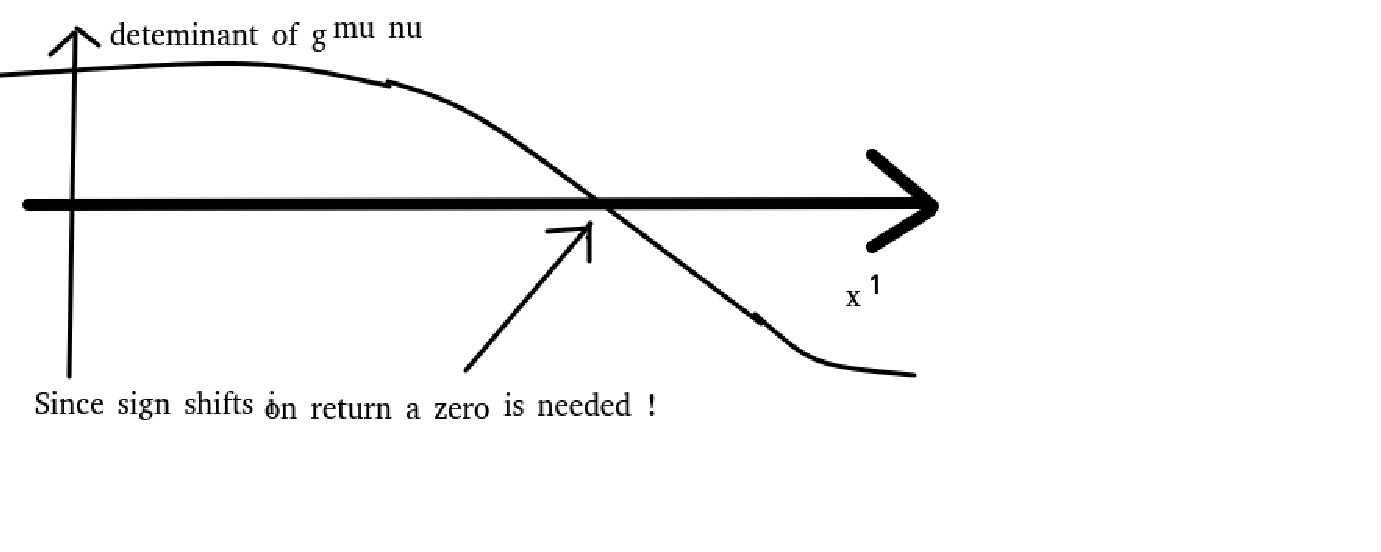}
  
\end{frame}

\begin{frame}
  {\bf We really needed upper index $g^{\mu\nu}$, so it must be ``fundamentally'' an effective (?) field}

  But the lower index ones $g_{\mu\nu}$ could just a definiton of an inverse.

  I.e. the $g_{\mu\nu}$ with lower indices would just be the defined as the
  inverse
  \begin{eqnarray}
    g_{\mu\nu}&=& (-1)^{\mu+\nu}\frac{det g^{..}|_{\hbox{left out}\mu\nu}}{det(g^{..})}
  \end{eqnarray}

  {\bf So when $det(g^{\mu\nu}) = 0$ (genericly) all matrix elements of
    $g_{\mu\nu}$ go to infinity.} And so near by all distances between the
  point in the projective space become huge.
\end{frame}

\section{Char}
\label{sChar}
\begin{frame}
  {\bf Characterization of Projective line as 3-transitive}

  In last years Bled talk I presented a work with Masao
  Ninomiya\cite{projectiveline}, in which we showed that requiring for
  a group acting on space $X$ in sharply 3-transitive way, essentially
  led you to the projective line (= a one dimensional projective space.

  {\bf Hope to somehow characterize projevtive spaces by some form of
    $n$-transitivity} (may be next years talk?) 
  \end{frame}
\begin{frame}
  {\bf Projective line}

  The projective line is the real axis extended with one point at infinity.
\end{frame}
\begin{frame}
  {\bf Projective space of $d$ dimension  as set of Rays in Vector space
    of $d+1$}
  
  \includegraphics[scale=0.7]{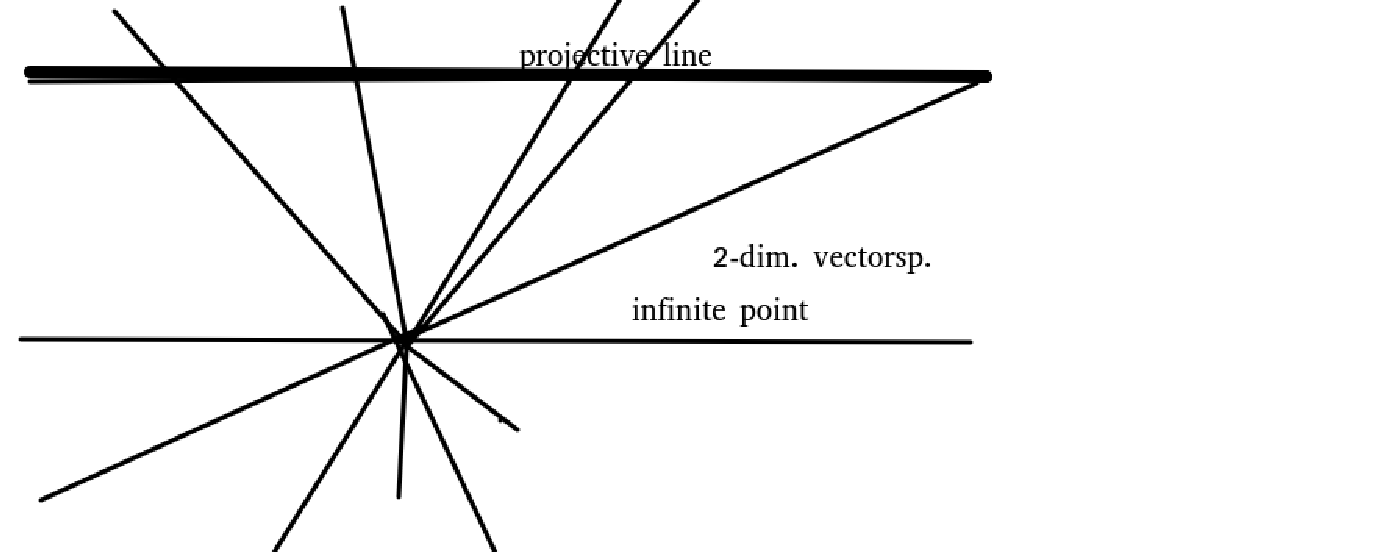}
  \end{frame}

\section{Concl.}
\label{sC}
\begin{frame}
  {\bf Conclusion}
  \begin{itemize}
  \item We centered the present talk about:
    That Astri Kleppe and I could derive : {\bf Principle of locality, that
      the action effectively is an integral over a local Lagrange density
      - only depending on fields defined at a single space-time point -.}

    From: {\bf Diffeomorphism symmetry i.e. a manifold} and some milder
    assumptions, Taylor expandability, keeping only low dimensional terms.
  \end{itemize}
\end{frame}
\begin{frame}
  {\bf Concluison (continued)}
The locality-derivation were not quite successful:
  \begin{itemize}
  \item A couple of not quite succeedings turned out promissing:
    \begin{itemize}
    \item Only getting a function form $S[\phi]=F(S_1[\phi],...,S_n[\phi])$,
      where the $S_i[\phi]$ are truly local actions, meant: couplings could
      depend on what goes on all over at all times, we could likely get our
      several vacua with same energy density.

    \item Would get superlocality and thus no propagation, unless we have
      some spontaneous breaking of diffeomorphism symmetry down to a metric
      space time, by {\bf gravity fields}
    \end{itemize}
  \item A phenomenological argument - based on only 2 s.d. though - for
    imbedding of world in a {\bf projective space-time}.
    \end{itemize}
  
  \end{frame}
\begin{frame}
  {\bf Conclusion (a bit old)}
  \begin{itemize}
  \item Screwed logic: If we want to use Astris and my way to derive
    locallity, then space-time must be diffeomorphism invariant, or it might
    still go with less symmetry, such as the {\bf Projective space time}.
  \item With such diffeoemorphism or projective space symmetry, there would
    be no propagations of signals no waves, if there does not appear by
    spontaneous break down a $g^{\mu\nu}$ (with upper indices) being non-zero.
  \item So gravity is activity due to a needed for propagation, non-zero
    field. 
    \end{itemize}
  
\end{frame}
\begin{frame}
  {\bf Speculative Conclusion}

  Fundamentally we have for some reason a projective space or a manifold with
  diffeomorphism invariance - in any case a space-time with symmetry group
  acting in a practically $n$-transitive way with a high $n$ - but then
  either a field $g^{\mu\nu}(x)$ or some corresponding vierbein fields
  $V^{\mu}_a(x)$ (also with upper curved indices) get non-zero in the vacuum.
  This makes possible propagation of waves/particles along the direction of
  the subspace of the tangent space  spanned by this  $g^{\mu\nu}(x)$ or these
  vierbeins  $V^{\mu}_a(x)$. So at the end the end {\bf the Einstein general
    relativity four-space is imbedded into the more fundamental general
    manifold or projective space.} 
\end{frame}

\begin{frame}
  {\bf Aacknowledgement:}

  I thank Danai Roumelioti for calling my attention to the works by
  R. Percacci \cite{Percacci}, and her and George Zoupanos for discussions.
  Also I thnk the Niels Bohr Institute for staying there as emeritus and for
  support to the Bled-Workshop, where I also talked about this subject. Most
  of the work presented here was a review of my work with Astri Kleppe, with
  whom I had very many discussions.
\end{frame}


\begin{thebibliography}{99}
  \bibitem{Percacci}
    Roberto Percacci, ``spontaneous Soldering'', Physics Letters vol. {\bf 144}
    number 1.2, 23. august 1984.

    R. Percacci, ``The Higgs phenomenon in Quantum Gravity'',
    arXiv: 0712.3545v2 [hep-th] 8. jan. 2008.

  
  \bibitem{locality}
H. B. Nielsen and A. Kleppe 
(The Niels Bohr Institute, Copenhagen, Denmark and 
SACT, Oslo, Norway)

``Towards a Derivation of Space''
arXiv: 1403.1410v1 [hep-th] 6 Mar 2014.
\bibitem{locality2}
H.B. Nielsen and A. Kleppe
BLED WORKSHOPS
IN PHYSICS
VOL. 20, NO. 2
Proceedings to the 22nd Workshop
What Comes Beyond . . . (p. 135)
Bled, Slovenia, July 6–14, 2019

`` Deriving Locality From Diffeomorphism
Symmetry in a Fiber Bundle Formalism ''


\bibitem{symp}C. Godbillon, "Géométrie différentielle et mécanique
  analytique" , Hermann (1969).
  
\bibitem{Paston}
A.A. Sheykin and S.A. Paston
The approach to gravity as a theory of embedded surface
AIP Conference Proceedings 1606, 400–406 (2014)

\bibitem{Smallrep}
Holger Bech Nielsen,
Niels Bohr Institutet, Blegdamsvej 15 -21 DK 2100Copenhagen
``Small Representation Explaing, Why Standard
Model Group'',
PoS(CORFU2014)045 (Corfu 2014).
\bibitem{Smallrep2}H. B. Nielsen, “Dimension Four Wins the Same Game as the
  Standard Model Group,”
  arXiv:1304.6051 [hep-ph].
\bibitem{SmallrepDon}  
Don Bennett and H. B. Nielsen, “Seeking...”, Contribution to the workshop
“Beyond the Standard
Models”, Bled 2011
\bibitem{Smallrep3} H. B. Nielsen, “Small Representation Principle”,
  Contribution to the workshop “Beyond the Standard
Models”, Bled 2013.
\bibitem{ORaifeartaigh}
O’Raifeartaigh, Group Structure of Gauge theories,University Press Cambridge (1986)
\bibitem{WMAPlowl}
P. Bielewicz, K. M. Górski, A. J. Banday
``Low-order multipole maps of cosmic microwave background anisotropy
derived from WMAP'',
Monthly Notices of the Royal Astronomical Society, Volume 355, Issue 4,
December 2004, Pages 1283–1302,
https://doi.org/10.1111/j.1365-2966.2004.08405.x



\bibitem{Stillits}
Stillits, Thesis for Cand.Scient.degree at The Niels Bohr Insitute.  
\bibitem{projectiveline}
  H.B. Nielsen, Masao Ninomiy ``String Field Theory and Bound State,
  Projective Line, and sharply 3-transitive group''
  Proceedings contribution to the 24th Bled-Workshop "What comes
  beyond the Standard Models" Bled, July 3-11, 2021, Slovenia;
 	arXiv:2111.05106 [physics.gen-ph]
  	(or arXiv:2111.05106v1 [physics.gen-ph] for this version)
  	
https://doi.org/10.48550/arXiv.2111.05106
\bibitem{Simonov}
A. A. Simonov, On generalized sharply n-transitive groups,
Izvestiya: Mathematics, 2014, Volume 78, Issue 6, 1207–
1231
Math-Net.Ru
  ``General Russian Mathematical Portal''

DOI: 10.1070/IM2014v078n06ABEH002727

    \bibitem{Explaining}
Mayukh R. Gangopadhyaya, Grant J. Mathews, Kiyotomo Ichiki, Toshitaka
``Explaining low ` Anomalies in the CMB Power Spectrum with Resonant
Superstring Excitations during Inflation''
Eur. Phys.J., arXiv:1701.00577v5[astro-ph.CO] 17 sep. 2018
\vspace{-2mm}
\bibitem{punctuated}
Mussadiq H. Qureshi, Asif Iqbal, Manzoor A. Malik, and Tarun Souradeep
``Low-l power suppression in punctuated inflation''
Published 7 April 2017; 2017 IOP Publishing Ltd and Sissa Medialab srl
Journal of Cosmology and Astroparticle Physics, Volume 2017, April 2017
Citation Mussadiq H. Qureshi et al JCAP04(2017)013,
DOI 10.1088/1475-7516/2017/04/013 
 




  \end{thebibliography}
\end{document}